\documentclass[12pt]{article} 
\usepackage[utf8]{inputenc} 
\usepackage{geometry}
\usepackage{graphicx} 
\usepackage{amssymb}
\usepackage{amsmath}
\usepackage{cite}
\usepackage{mathrsfs}
\usepackage{booktabs} 
\usepackage{array} 
\usepackage{paralist}
\usepackage{verbatim} 
\usepackage{subfig}
\usepackage{fancyhdr} 
\usepackage{sectsty}
\allsectionsfont{\sffamily\mdseries\upshape} 
\usepackage[nottoc,notlof,notlot]{tocbibind} 
\usepackage[titles,subfigure]{tocloft} 

\title{Polynomial solutions of the boundary value problems for the Poisson equation in a layer.}
\author{\bf Oleg D. Algazin}
\date{Bauman Moscow State Technical University, Moscow, Russia} 

\begin{document}
\maketitle
\thispagestyle{empty}

\begin{abstract}
     In a multidimensional infinite layer bounded by two hyperplanes, the Poisson equation with the polynomial right-hand side is considered. It is shown that the Dirichlet boundary value problem and the mixed Dirichlet-Neumann boundary value problem with polynomial boundary conditions have a unique solution in the class of functions of polynomial growth and it solution is a polynomial. An algorithm for constructing this polynomial solution is given and examples are considered.
\end{abstract}


\section*{Introduction}

      It is well known that the Dirichlet problem for the Laplace equation in a ball has a unique polynomial solution (harmonic polynomial) in the case if the given boundary value is the trace of an arbitrary polynomial on the sphere \cite{S},\cite{SW}. S.M.Nikol'skii \cite{N1} generalized this result to the case of a boundary value problem of the first kind for a linear differential self-adjoint operator of order $2l$ with constant coefficients (in particular polyharmonic) and for a domain that is an ellipsoid in $\mathbb{R}^n$. The corresponding inhomogeneous equation was considered in \cite{N2}. For a polyharmonic equation in a ball (homogeneous and inhomogeneous), the algorithm for constructing a polynomial solution of the Dirichlet problem on the basis of the Almansi formula was proposed by V.V. Karachik \cite{K}.

     S.M.Nikol'skii showed in \cite{N1} that for a domain bounded by an algebraic surface of elliptic type of degree greater than two or bounded by several pieces of algebraic surfaces, the above statement is false, that is, not for any polynomial boundary value, the solution is a polynomial. For example, in the two-dimensional case for the Laplace operator and for a region that is a polygon. Criteria for the solvability of boundary value problems for the Laplace and Poisson equations in polynomials for polygons are considered in the papers of E.A.Volkov \cite{V1},\cite{V2}.

      In this paper we consider the Poisson equation in an unbounded domain (in a layer) with a polynomial right-hand side
\[
\Delta u(x,y)=P(x,y),~~   x\in\mathbb{R}^n,~~   0<y<a, \eqno{(1)}
\]
where $x=(x_1,…,x_n ),~\Delta$ is the Laplace operator,
\[
\Delta=\frac{\partial^2 }{\partial x_1^2}+...+\frac{\partial^2 }{\partial x_n^2}+\frac{\partial^2 }{\partial y^2},
\]
$P(x,y)$ is a  polynomial in the variables $x$ and $y$.

     On the boundary of the layer we set the Dirichlet boundary conditions
\[
u(x,0)=\phi(x),~~u(x,a)=\psi(x),~~x\in\mathbb{R}^n, \eqno{(2)}
\]
and mixed Dirichlet-Neumann boundary conditions
\[
u(x,0)=\phi_1(x),~~u_y(x,a)=\psi_1(x),~~x\in\mathbb{R}^n, \eqno{(3)}
\]
where $\phi(x),\psi(x),\phi_1 (x),\psi_1 (x)$ are polynomials.

If $\tilde u(x,y)$ is a particular polynomial solution of the Poisson equation (1), then for the function $v(x,y)=u(x,y)-\tilde u(x,y)$, we obtain the Laplace equation
\[
\Delta v(x,y)=0,~~   x\in\mathbb{R}^n,~~   0<y<a, \eqno{(4)}
\]
and the Dirichlet boundary conditions
\[
v(x,0)=\phi(x)-\tilde u(x,0),~~v(x,a)=\psi(x)-\tilde u(x,a),~~x\in\mathbb{R}^n, \eqno{(5)}
\]
and mixed Dirichlet-Neumann boundary conditions
\[
v(x,0)=\phi_1(x)-\tilde u(x,0),~~v_y(x,a)=\psi_1(x)-\tilde u_y(x,a),~~x\in\mathbb{R}^n, \eqno{(6)}
\]
Solving boundary value problems for the Laplace equation (4), (5), and (4), (6), we obtain solutions of the boundary value problems for the Poisson equation (1), (2) and (1), (3) according to the formula
\[
u(x,y)=v(x,y)+\tilde u(x,y).
\]
If solutions of boundary value problems are sought in the class of functions of polynomial growth with respect to $x$:
\[
\int_{\mathbb{R}^n}|u(x,y)|(1+|x|)^{-m}dx<C,~~|x|=\sqrt{x_1^2+...+x_n^2}\eqno{(7)}
\]
for some $m\ge0$ and for each $y\in(0,a)$, then the solution is unique. The solutions of  boundary value problems for the Laplace equation are written in the form of convolutions of boundary functions with the corresponding kernels, which are fundamental solutions of the corresponding boundary value problems [8], [9]. It is shown that these solutions are polynomials and  algorithm for obtaining them is given. The formula for obtaining a particular polynomial solution for the Poisson equation is also given and examples are considered.

\section{A particular solution of the Poisson equation.  }
The Poisson equation (1) with the polynomial right-hand side $P(x,y)$,
\[
\Delta u(x,y)=P(x,y),~~   x\in\mathbb{R}^n,~~   0<y<a, 
\]
has polynomial solutions, one of which can be obtained from the formula given below. It is sufficient to give this formula for a monomial.

We first consider the case $n=1,~P(x,y)=x^k y^m$. Then a particular solution of the Poisson equation with the right-hand side $x^k y^m$ is the function
\[
\tilde u(x,t)=\Delta^{-1}(x^ky^m)=
\]
\[
=\sum_{j=0}^{[k/2]}(-1)^j\frac{k!m!}{(m+2j+2)!(k-2j)!}y^{m+2j+2}x^{k-2j},\eqno{(8)}
\]
where $[k/2]$ is the integer part of the number $k/2$.

By interchanging $x$ and $y$, the particular solution can be obtained in the form
\[
\tilde u_1(x,t)=\Delta^{-1}(y^mx^k)=
\]
\[
=\sum_{j=0}^{[m/2]}(-1)^j\frac{k!m!}{(k+2j+2)!(m-2j)!}x^{k+2j+2}y^{m-2j},\eqno{(9)}
\]
The degrees of the polynomials $\tilde u(x,y)$ and $\tilde u_1 (x,y)$ are two more than degree of the $P(x,y)$. 

The validity of formulas (8) and (9) is proved by direct verification
\[
\Delta \tilde u(x,y)=\Delta \tilde u_1(x,y)=x^ky^m.
\]
For example, for
\[
P(x,y)=x^4y^3
\]
a particular solution according to formula (8) will be
\[
\tilde u(x,y)=\frac{1}{20}x^4y^5-\frac{1}{70}x^2y^7+\frac{1}{2520}y^9,
\]
and according to formula (9) 
\[
\tilde u_1(x,y)=\frac{1}{30}x^6y^3-\frac{1}{280}x^8y.
\]

We now consider the case $n>1$ and the monomial
\[
P(x,y)=x^ky^m,
\]
where
\[
x^k=x_1^{k_1}... x_n^{k_n},~~k=(k_1,... ,k_n) ~\text{is multiindex,}
\]
\[
|k|=k_1+...+k_n~ \text{ is the degree of the monomial}~ x^k.
\]
A particular solution of the Poisson equation with the right-hand side $P(x,y)$ can be obtained according to formula
\[
\tilde u(x,t)=\Delta^{-1}(x^ky^m)=
\]
\[
=\sum_{j=0}^{[|k|/2]}(-1)^j\frac{m!}{(m+2j+2)!}y^{m+2j+2}\Delta_x^jx^k,\eqno{(10)}
\]
where
\[
\Delta_x^j=\left(\frac{\partial^2}{\partial x_1^2}+...+\frac{\partial^2}{\partial x_n^2}\right)^j.
\]
For  $n=1$, formula (8) is obtained from formula (10). Formula (10) is proved by direct verification. Changing the places $y$ with $x_j,j=1,...,n$, one can obtain $n$ more formulas for particular solutions of the Poisson equation with the right-hand side  $x^k y^m$.

For example, for
\[
P(x,y)=x^{(3,2,1)}y^3=x_1^3x_2^2x_3y^3
\]
a particular solution according to formula (10) will be
\[
\tilde u(x,y)=\frac{1}{20}x_1^3x_2^2x_3y^5-\frac{1}{420}x_1^3x_3y^7-\frac{1}{140}x_1x_2^2x_3y^7+\frac{1}{2520}x_1x_3y^9.
\]
The degree of the polynomial $\tilde u(x,y)$ are two more than degree of the $P(x,y)$.

\section{Uniqueness of the solution of boundary value problems}
We show that in the class of  functions of slow growth with respect to $x$, satisfying condition (7), the homogeneous equation (Laplace equation ) with homogeneous boundary conditions has only the trivial solution.

Since the functions $u (x,y)$ of slow growth with respect to $x$ for each $y\in (0,a)$ determine  regular functionals from the space  of generalized functions of slow growth $\mathscr{S}'(\mathbb{R}^n  )$ , we can apply  the Fourier transform to this functions with respect to $x$ [10]:
\[
\mathscr{F}_x [u(x,y) ](t,y)=U(t,y).
\]
We apply the Fourier transform with respect to $x$ to Laplace equation
\[
\Delta u(x,y)=0,~~x\in\mathbb{R}^n,~~0<y<a, 
\]
and to the homogeneous Dirichlet boundary conditions
\[
u(x,0)=0,~~u(x,a)=0
\]
and to the homogeneous Dirichlet-Neumann boundary conditions
\[
u(x,0)=0,~~u_y (x,a)=0.
\]
We obtain an ordinary second-order differential equation with parameter $t\in\mathbb{R}^n$
\[
-|t|^2 U(t,y)+U_{yy} (t,y)=0,~~0<y<a     \eqno{(11)}
\]
and boundary conditions
\[
U(t,0)=0,~~U(t,a)=0                              \eqno{(12)}
\]
or
\[
U(t,0)=0,~~U_y (t,a)=0.                          \eqno{(13)}
\]
The general solution of equation (11) has the form
\[
U(t,y)=c_1(t)\cosh(|t|y)+c_2(t)\sinh(|t|y).     \eqno{(14)}
\]
Substituting the boundary conditions (12) into (14), we obtain
\[
c_1(t)=0,~~c_2(t)\sinh(|t|a)=0.
\]
The last equation is equivalent to equation
\[
|t|c_2 (t)=0
\]
and the solution of the boundary value problem
\[
U(t,y)=c_1(t)\cosh(|t|y)+c_2(t)\sinh(|t|y)=c_2(t)\sinh(|t|y)=
\]
\[
=c_2(t)|t|\left(y+\frac{|t|^2y^3}{3!}+\frac{|t|^4y^5}{5!}+\dots\right)=0.
\]
Substituting the boundary conditions (13) into (14), we obtain
\[
c_1(t)=0,~~c_2(t)|t|\cosh(|t|a)=0.
\]
The last equation is equivalent to equation
\[
|t|c_2 (t)=0
\]
and, consequently,
\[
U(t,y)=0.
\]

We now consider polynomial solutions of boundary value problems for the Laplace equation, to which the solutions of boundary value problems for the Poisson equation with the polynomial right-hand side are reduced.

\section{Solutions of boundary value problems for the Laplace equation}

We consider the Laplace equation
\[
\Delta u(x,y)=0,~~   x\in\mathbb{R}^n,~~   0<y<a, \eqno{(15)}
\]
 Dirichlet boundary conditions
\[
u(x,0)=\phi(x),~~u(x,a)=\psi(x),~~x\in\mathbb{R}^n, \eqno{(16)}
\]
and Dirichlet-Neumann boundary conditions
\[
u(x,0)=\phi_1(x),~~u_y(x,a)=\psi_1(x),~~x\in\mathbb{R}^n, \eqno{(17)}
\]
where $\phi(x),\psi(x),\phi_1 (x),\psi_1 (x)$ are polynomials.

    It was shown in [8], [9] that if $\phi(x),\psi(x),\phi_1(x),\psi_1(x)$ are generalized functions of slow growth (in particular, polynomials), then the solutions of these problems can be written in the form of the convolution of boundary functions with kernels that are fundamental solutions of these boundary value problems. The solution of the Dirichlet problem (15), (16) is written in the form
\[
u(x,y)=\phi(x)*P_n(|x|,a-y)+\psi(x)*P_n(|x|,y), \eqno{(18)}
\]   
and solution of the Dirichlet-Neumann problem (15), (17) is written in the form
\[
u(x,y)=\phi_1(x)*K_n(|x|,y)+\psi_1(x)*L_n(|x|,y).     \eqno{(19)}
\]
The kernels $P_n(|x|,a-y),P_n(|x|,y),K_n(|x|,y),L_n(|x|,y)$ are fundamental solutions of boundary value problems, that is, these kernels are solutions of the Laplace equation (15), which satisfy the boundary conditions, respectively
\[
u(x,0)=\delta(x),~~u(x,a)=0,~~x\in\mathbb{R}^n, \eqno{(20)}
\]
\[
u(x,0)=0,~~u(x,a)=\delta(x),~~x\in\mathbb{R}^n, \eqno{(21)}
\]
\[
u(x,0)=\delta(x),~~u_y(x,a)=0,~~x\in\mathbb{R}^n, \eqno{(22)}
\]
\[
u(x,0)=0,~~u_y(x,a)=\delta(x),~~x\in\mathbb{R}^n, \eqno{(23)}
\]
where $\delta(x)$ is Dirac $\delta$-function.

The solutions (18), (19) are unique in the class of functions of slow growth with respect to $x$. We show that they are polynomials and give an algorithm for their finding.

\subsection{The Dirichlet problem}

We apply the Fourier transform with respect to $x$ to the Laplace equation (15) and the boundary conditions (21), we obtain the boundary value problem for the second order ODE (11) with boundary conditions
\[
U(t,0)=0,~~U(t,a)=1.                       \eqno{(24)}
\]       
Its solution is the function
\[
U(t,y)=p_n(|t|,y)=\frac{\sinh(|t|y)}{\sinh(|t|a)}.\eqno{(25)}
\]     
In the case of boundary conditions (20) we obtain the function $p_n(| t |,a-y)$. We apply the inverse Fourier transform to (25), we obtain a fundamental solution of the Dirichlet problem (the Poisson kernel)
\[
P_n(|x|,y)=\mathscr{F}_t^{-1}\left[\frac{\sinh(|t|y)}{\sinh(|t|a)}\right](x,y)=
\frac{1}{(2\pi)^n}\int_{\mathbb{R}^n}\frac{\sinh(|t|y)}{\sinh(|t|a)}e^{-ixt}dt,\eqno{(26)}
\]
where $tx = t_1x_1+t_2x_2+...+t_nx_n$. Passing to spherical coordinates and denoting $|x|=r,|t|=\rho$, we obtain
\[
P(r,y)=\frac{1}{(2\pi)^{n/2}r^{n/2-1}}\int_0^{\infty}\frac{\sinh(\rho y)}{\sinh(\rho a)}\rho^{n/2}J_{n/2-1}(r\rho),\eqno{(27)}
\]
where $J_{n/2-1}(r\rho)$ is the Bessel function of the first kind of the order $\nu=n/2-1$. This formula is also valid for $n=1$. For $n=1$, the integral (27) is calculated in an explicit form [9]
\[
P_1(x,y)=P_1 (|x|,y)=\frac{1}{2a}\frac{\sin(\pi y/a)}{\cosh(\pi x/a)+\cos(\pi y/a)},\eqno{(28)}
\]
\[
P_1(x,a-y)=\frac{1}{2a}\frac{\sin(\pi y/a)}{\cosh(\pi x/a)-\cos(\pi y/a)}.
\]
This formula is well known [11, c.187], [12]. In the case of arbitrary dimension $n$, the Poisson kernel in the form (25) was considered in [13]. In [9] a recurrence formula with respect to dimension is received
\[
P_{n+2}(r,y)=-\frac{1}{2\pi r}\frac{\partial}{\partial r}P_n(r,y).\eqno{(29)}
\]
According to this formula, for example,
\[
P_3(|x|,y)=\frac{1}{4a^2}\frac{\sin(\pi y/a)\sinh(\pi |x|/a)}{|x|\left(\cosh(\pi |x|/a)+\cos(\pi y/a)\right)^2}.\eqno{(30)}
\]

The solution of the Dirichlet problem is given by the formula (18), where $\phi(x),\psi(x)$ are polynomials. Let $\phi(x)=0$, then
\[
u(x,y)=\psi(x)*P_n(|x|,y)=\int_{\mathbb{R}^n}\psi(t)P_n(|x-t|,y)dt=\int_{\mathbb{R}^n}\psi(x-t)P_n(|t|,y)dt.\eqno{(31)}
\]
We show that this integral is a polynomial and we will give an algorithm for obtaining it. For this we do not need the explicit form of the kernel $P_n(| x |,y)$. To prove that the integral (31) is a polynomial with respect to $(x,y)$ it is sufficient to prove this for the monomial $\psi(t)=t^k$.

\subsubsection{The case $n=1$}

Let
\[
\psi(t)=t^0=1.
\]
The solution of the Dirichlet problem with boundary conditions
\[u(x,0)=0,~~u(x,a)=1,~~x\in\mathbb{R},
\]
will be a function
\begin{gather*}
u_0(x,y)=\int_{-\infty}^{\infty}P_1(t,y)dt=\lim_{x\to 0}\int_{-\infty}^{\infty}P_1(t,y)e^{ixt}dt=\\
=\lim_{x\to 0}\mathscr{F}_t\left[P_1(t,y)\right](x,y)=\lim_{x\to 0}\frac{\sinh(xy)}{\sinh(xa)}=\frac{y}{a}.
\end{gather*}
Let now
\[
\psi(t)=t^k.
\]
The solution of the Dirichlet problem with boundary conditions
\[
u(x,0)=0,~~u(x,a)=x^k,~~x\in\mathbb{R},
\]
will be a function
\begin{gather*}
u_k(x,y)=\int_{-\infty}^{\infty}(x-t)^kP_1(t,y)dt=\\
=\int_{-\infty}^{\infty}\sum_{j=0}^kC_k^jx^{k-j}t^j(-1)^jP_1(t,y)dt=\sum_{j=0}^kC_k^jx^{k-j}(-1)^j\int_{-\infty}^{\infty}t^jP_1(t,y)dt.
\end{gather*}
where $C_k^j=k!/j!(k-j)!$ are binomial coefficients. Since the last integral for odd $j$ is zero because of the parity of $P_1(t,y)$ with respect to $t$, then
\[
u_k(x,y)=\sum_{m=0}^{[k/2]}C_k^{2m}x^{k-2m}\int_{-\infty}^{\infty}t^{2m}P_1(t,y)dt,
\]
where $[k/2]$ is the integer part of the number $k/2$. Using the properties of the Fourier transform, we obtain
\begin{gather*}
f_{2m}(y)=\int_{-\infty}^{\infty}t^{2m}P_1(t,y)dt=\lim_{x\to 0}\int_{-\infty}^{\infty}t^{2m}P_1(t,y)e^{ixt}dt=\\
=\lim_{x\to 0}\mathscr{F}_t\left[t^{2m}P_1(t,y)\right](x,y)=(-1)^m\lim_{x\to 0}\frac{d^{2m}}{dx^{2m}}\frac{\sinh(xy)}{\sinh(xa)}.
\end{gather*}
We expand in the series with respect to $x$ the even function
\[
\frac{\sinh(xy)}{\sinh(xa)}=c_0+c_2x^2+...+c_{2m}x^{2m}+...~.
\]
Performing the division of the series, we obtain recurrence formulas for the coefficients $c_{2m}$ 
\begin{gather*}
c_0=\frac{y}{a}\\
c_2=\frac{1}{a}\left(\frac{y^3}{3!}-\frac{a^3}{3!}c_0\right)=\frac{1}{3!a}(y^3-ya^2),...,
\end{gather*}
\[
c_{2m}=\frac{1}{a}\left(\frac{y^{2m+1}}{(2m+1)!}-\frac{a^3}{3!}c_{2m-2}-\frac{a^5}{5!}c_{2m-4}-...-\frac{a^{2m+1}}{(2m+1)!}c_0\right).\eqno{(32)}
\]
From these formulas it follows that $c_{2m}$ are polynomials of $y$ and, consequently,
\[
f_{2m}(y)=(-1)^m(2m)!c_{2m}
\]
are polynomials of $y$ and the solution of the problem
\[
u_k(x,y)=\sum_{m=0}^{[k/2]}C_k^{2m}x^{k-2m}f_{2m}(y)
\]
is a polynomial in $x$ and $y$.

We write out the first few polynomials $f_{2m}(y)$ and few solutions $u_k(x,y)$.
\begin{gather*}
f_0(y)=\frac{y}{a},\\
f_2(y)=-\frac{y}{3a}(y^2-a^2),\\
f_4(y)=\frac{y}{15a}(3y^4-10y^2 a^2+7a^4 ), \\
f_6(y)=-\frac{y}{21a}(3y^6-21y^4 a^2+49y^2 a^4-31a^6 ),\\
f_8(y)=\frac{y}{45a}(5y^8-60y^6 a^2+249y^4 a^4-620y^2 a^6+381a^8 ),\\
f_{10}(y)=-\frac{y}{33a}(3y^{10}-55y^8 a^2+462y^6 a^4-2046y^4 a^6+4191y^2 a^8-2555a^{10}).\\
u_0(x,y)=\frac{y}{a}  ,\\
u_1(x,y)=\frac{xy}{a},\\
u_2(x,y)=\frac{y}{3a}(3x^2-y^2+a^2 ),\\
u_3(x,y)= \frac{xy}{a}(x^2-y^2+a^2 ),\\
u_4(x,y)=\frac{y}{15a}(15x^4-30x^2 y^2+30x^2 a^2+3y^4-10y^2 a^2+7a^4 ),\\
u_5(x,y)=\frac{xy}{3a}(3x^4-10x^2 y^2+10x^2 a^2+3y^4-10y^2 a^2+7a^4 ).
\end{gather*}
The functions $v_k(x,y)=u_k(x,a-y)$ are solutions of the Laplace equation and satisfy the boundary conditions
\[
v_k(x,0)=x^k,~~   v_k (x,a)=0,~~x\in\mathbb{R}.
\]

If the functions $\phi(x)$ and $\psi(x)$ in the boundary conditions are arbitrary polynomials
\[
\phi(x)=a_0+a_1 x+a_2 x^2+...+a_k x^k,
\]
\[
\psi(x)=b_0+b_1 x+b_2 x^2+...+b_l x^l,
\]
then the solution of the boundary value problem is a polynomial
\[
u(x,y)=a_0 u_0 (x,a-y)+a_1 u_1 (x,a-y)+...+a_k u_k (x,a-y)+
\]
\[
+b_0 u_0 (x,y)+b_1 u_1 (x,y)+...+b_l u_l (x,y).
\]
\textbf{Remark 1. } 
The polynomials $f_{2m}(y)$ are the exact values of the integrals
\[
\frac{1}{2a}\int_{-\infty}^{\infty}\frac{x^{2m}\sin(\pi y/a)}{\cosh(\pi x/a)+\cos(\pi y/a)}dx=f_{2m}(y),~~-a<y<a.\eqno{(33)}
\]
for example,
\[
\frac{1}{2a}\int_{-\infty}^{\infty}\frac{x^{4}\sin(\pi y/a)}{\cosh(\pi x/a)+\cos(\pi y/a)}dx=\frac{y}{15a}(3y^4-10y^2 a^2+7a^4 ).
\]
Taking into account that
\[
\frac{1}{2a}\frac{x^{2m}\sin(\pi y/a)}{\cosh(\pi x/a)+\cos(\pi y/a)}=\text{Im}\left\{\frac{1}{a}\frac{x^{2m}}{e^{\pi x/a}e^{-i\pi y/a}+1}\right\},
\]
we express the integral (33) in terms of the polylogarithm
\[
\frac{1}{2a}\int_{-\infty}^{\infty}\frac{x^{2m}\sin(\pi y/a)}{\cosh(\pi x/a)+\cos(\pi y/a)}dx=
\]
\[
=-\frac{2(2m)!a^{2m}}{\pi^{2m+1}}\text{Im}\left\{\text{Li}_{2m+1}\left(-e^{i\pi y/a}\right)\right\}=\frac{2(2m)!a^{2m}}{\pi^{2m+1}}\sum_{k=1}^{\infty}(-1)^{k-1}\frac{\sin(\pi ky/a)}{k^{2m+1}}
\]
and we obtain the summation formula for the trigonometric series
\[
\sum_{k=1}^{\infty}(-1)^{k-1}\frac{\sin(\pi ky/a)}{k^{2m+1}}=\frac{\pi^{2m+1}}{2(2m)!a^{2m}}f_{2m}(y),~~-a<y<a.\eqno{(34)}
\]
For example, for $m=2,a=\pi$
\[
\sum_{k=1}^{\infty}(-1)^{k-1}\frac{\sin(ky)}{k^{5}}=\frac{y}{720}\left(3y^4-10\pi^2y^2+7\pi^4\right),~~-\pi<y<\pi.
\]
Similarly we obtain formulas
\[
\frac{1}{2a}\int_{-\infty}^{\infty}\frac{x^{2m}\sin(\pi y/a)}{\cosh(\pi x/a)-\cos(\pi y/a)}dx=f_{2m}(a-y),
\]
\[
\sum_{k=1}^{\infty}\frac{\sin(\pi ky/a)}{k^{2m+1}}=\frac{\pi^{2m+1}}{2(2m)!a^{2m}}f_{2m}(a-y),~~0<y<2a.\eqno{(35)}
\]
For example
\[
\sum_{k=1}^{\infty}\frac{\sin(ky)}{k^{5}}=\frac{y(\pi-y)}{720}(3y^3-12\pi y^2+8\pi^2y+8\pi^3),~~0<y<2\pi.
\]
In the handbook [14, p.726] the sums of the series (34) and (35) are expressed in terms of the Bernoulli polynomials.

\textbf{Example 1. }
Consider the Dirichlet problem for the Poisson equation
\[
\Delta u(x,y)=x^4 y^3,~~-\infty<x<\infty,~~0<y<1,
\]
\[
u(x,0)=8x^4-8x^2+1,~~u(x,1)=8x^4-8x^2+1,~~-\infty<x<\infty.
\]
A particular solution of the Poisson equation is the function
\[
\tilde u(x,y)=\frac{1}{20} x^4 y^5-\frac{1}{70} x^2 y^7+\frac{1}{2520} y^9
\]
and the Dirichlet problem for the Poisson equation reduces to the Dirichlet problem for the Laplace equation for the function $v(x,y)=u(x,y)–\tilde u(x,y)$:
\begin{gather*}
\Delta v(x,y)=0,~~-\infty<x<\infty,~~0<y<1,\\
v(x,0)=u(x,0)-\tilde u(x,0)=8x^4-8x^2+1,\\
v(x,1)=u(x,1)-\tilde u(x,1)=\frac{159}{20} x^4-\frac{559}{70}x^2+\frac{2519}{2520}.
\end{gather*}
The solution of this problem is the function
\begin{gather*}
v(x,y)=8 u_4(x,a-y)-8 u_2 (x,a-y)+u_0(x,a-y)+\\
+\frac{159}{20}u_4(x,y)-\frac{559}{70}u_2(x,y)+\frac{2519}{2520}u_0(x,y),~~\text{where}~ a=1.
\end{gather*}
The solution of the original problem is the function
\begin{gather*}
u(x,y)=v(x,y)+\tilde u(x,y)=\\
=\frac{1}{20} x^4 y^5-\frac{1}{20} x^4 y+8x^4-\frac{1}{70} x^2 y^7+\frac{1}{10} x^2 y^3-48x^2 y^2+\frac{1677}{35} x^2 y-\\
-8x^2+{1}{2520} y^9-\frac{1}{100}y^5+8y^4-\frac{559}{35} y^3+8y^2-\frac{239}{12600} y+1.
\end{gather*}

\textbf{Remark 2. }
If we look for a polynomial solution of the Dirichlet problem in the rectangle $0\le x\le b,0\le y\le a$ with polynomial boundary conditions, then the setting of polynomials on two parallel sides of the rectangle uniquely determines the polynomial solution of the Dirichlet problem in the strip and, therefore, uniquely determines the polynomial boundary values on the other two sides. If other polynomials were initially defined on these sides, then there is no polynomial solution of the Dirichlet problem in a rectangle with these boundary values. For example, in the rectangle $0\le x\le 1,0\le y\le 1$ we define the boundary conditions $u(x,0) = x^2,~u (x,1) = x^2$ on two parallel sides, then the unique polynomial solution of the Dirichlet problem for the strip
\begin{gather*}
\Delta u(x,y)=0,~~-\infty<x<\infty,~~0<y<1,\\
u(x,0)=x^2,~~u(x,1)=x^2,
\end{gather*}
is a function
\[
u(x,y)=x^2-y^2+y,
\]
which on the other two sides takes the values
\[
u(0,y)=-y^2+y,~   u(1,y)=1-y^2+y.
\]
If you set other values on these sides, for example,
\[
u(0,y)=0,~   u(1,y)=1,
\]
then the unique solution of the Dirichlet problem for a rectangle  is not a polynomial.

\subsubsection{The case $n>1$}
If 
\[
\psi(t)=1,~~t\in\mathbb{R}^n,
\]
then the solution of the Dirichlet problem with the boundary condition
\[
u(x,0)=0,~~u(x,a)=1,~~x\in\mathbb{R}^n,
\]
is a function
\begin{gather*}
u_0(x,y)=\int_{\mathbb{R}^n}P_n(|t|,y)dt=\lim_{x\to 0}\int_{\mathbb{R}^n}P_n(|t|,y)e^{ixt}dt=\\
=\lim_{x\to 0}\mathscr{F}_t\left[P_n(|t|,y)\right](x,y)=\lim_{x\to 0}\frac{\sinh(|x|y)}{\sinh(|x|a)}=\frac{y}{a}.
\end{gather*}
If
\[
\psi(t)=t^k,~~t\in\mathbb{R}^n,~~   k~ \text{is multiindex},
\]
then the solution of the Dirichlet problem with the boundary condition
\[
u(x,0)=0,~~u(x,a)=x^k,~~x\in\mathbb{R}^n ,
\]
is a function
\[
u_k(x,y)=\int_{\mathbb{R}^n}(x-t)^kP_n(|t|,y)dt,
\]
where
\[
(x-t)^k=(x_1-t_1 )^{k_1}(x_2-t_2 )^{k_2}…(x_n-t_n )^{k_n}     \eqno{(36)}
\]
and this integral will be nonzero only for those monomials of the polynomial (36) that contain $t_j$  in even powers. Therefore
\[
u_k(x,y)=\sum_{m=0}^{[k/2]}C_k^{2m}x^{k-2m}\int_{\mathbb{R}^n}t^{2m}P_n(|t|,y)dt=\sum_{m=0}^{[k/2]}C_k^{2m}x^{k-2m}f_{2m}(y),
\]
where
\[
C_k^{2m}=C_{k_1}^{2m_1}C_{k_2}^{2m_2}...C_{k_n}^{2m_n},~~[k/2]=([k_1/2],[k_2/2],...,[k_n/2]).
\]
Using the properties of the Fourier transform [10], we obtain
\begin{gather*}
f_{2m}(y)=\int_{\mathbb{R}^n}t^{2m}P_n(|t|,y)dt=\lim_{x\to 0}\int_{\mathbb{R}^n}t^{2m}P_n(|t|,y)e^{ixt}dt=\\
=\lim_{x\to 0}\mathscr{F}_t\left[t^{2m}P_n(|t|,y)\right](x,y)=(-1)^{|m|}\lim_{x\to 0}\partial_x^{2m}\frac{\sinh(|x|y)}{\sinh(|x|a)},
\end{gather*}
where
\[
\partial_x^{2m}=\frac{\partial^{2|m|}}{\partial x_1^{2m_1}\partial x_2^{2m_2}...\partial x_n^{2m_n}}
\]
We have the expansion
\begin{gather*}
\frac{\sinh(|x|y)}{\sinh(|x|a)}=c_0+c_2|x|^2+...+c_{2|m|}|x|^{2|m|}+...=\\=c_0+c_2(x_1^2+...+x_n^2)+...+c_{2|m|}(x_1^2+...+x_n^2)^{|m|}+...~.
\end{gather*}
The coefficients $c_{2|m|}$ are found from formulas (32) and polynomials
\[
f_{2m}(y)=(-1)^{|m|}\frac{(2m)!|m|!}{m!}c_{2|m|}=\frac{(2m)!|m|!}{|2m|!m!}f_{2|m|}(y).
\]
For example,
\[
f_{2(2,1,1)}(y)=\frac{1}{35}f_8(y).
\]
This polynomial is the exact value of the integral
\begin{gather*}
\frac{1}{4a^2}\int_{\mathbb{R}^3}\frac{x^{(4,2,2)}\sin(\pi y/a)\sinh(\pi |x|/a)}{|x|\left(\cosh(\pi |x|/a)+\cos(\pi y/a)\right)^2}=\frac{1}{35}f_8(y)=\\
=\frac{y}{1575a}(5y^8-60y^6a^2+249y^4a^4-620y^2a^6+381a^8).
\end{gather*}
The functions $v_k(x,y)=u_k(x,a-y)$ are solutions of the Laplace equation and satisfy the boundary conditions
\[
v_k(x,0)=x^k,~~   v_k(x,a)=0,~~x\in\mathbb{R}^n,~~k~\text{ is multiindex} .
\]
\textbf{Example 2. }
We consider the Dirichlet problem for the Poisson equation
\begin{gather*}
\Delta u(x,y)=x^{(3,2,1)}y^3=x_1^3 x_2^2 x_3 y^3,~~x\in\mathbb{R}^3,~~0<y<1,\\
u(x,0)=0,~~u(x,1)=0,~~x\in\mathbb{R}^3.
\end{gather*}
A particular solution of the Poisson equation is the function
\[
\tilde u(x,y)=\frac{1}{20}x_1^3x_2^2x_3y^5-\frac{1}{420}x_1^3x_3y^7-\frac{1}{140}x_1x_2^2x_3y^7+\frac{1}{2520}x_1x_3y^9.
\]
and the Dirichlet problem for the Poisson equation reduces to the Dirichlet problem for the Laplace equation for the function $v(x,y)=u(x,y) –\tilde u(x,y)$:
\begin{gather*}
\Delta v(x,y)=0,~~x\in\mathbb{R}^3,~~0<y<1,\\
v(x,0)=u(x,0)-\tilde u(x,0)=0,\\
v(x,1)=u(x,1)-\tilde u(x,1)=\\
=-\frac{1}{20} x_1^3 x_2^2 x_3+\frac{1}{420} x_1^3 x_3+\frac{1}{140} x_1 x_2^2 x_3-\frac{1}{2520}x_1 x_3.
\end{gather*}
The solution of the Dirichlet problem for the Poisson equation is the function
\begin{gather*}
u(x,y)=\tilde u(x,y)+v(x,y)=\tilde u(x,y)-\\
-\frac{1}{20}u_{(3,2,1)}(x,y)+\frac{1}{420}u_{(3,0,1)}(x,y)+\frac{1}{140}u_{(1,2,1)} (x,y)-\frac{1}{2520}u_{(1,0,1)}(x,y)=\\
=\frac{1}{20} x_1^3 x_2^2 x_3 y^5-\frac{1}{20}x_1^3 x_2^2 x_3 y
-\frac{1}{420} x_1^3 x_3 y^7+\frac{1}{60} x_1^3 x_3 y^3-\frac{1}{70} x_1^3 x_3 y-\\
-\frac{1}{140}x_1 x_2^2 x_3 y^7+\frac{1}{20} x_1 x_2^2 x_3 y^3-\frac{3}{70} x_1 x_2^2 x_3 y+\frac{1}{2520}x_1 x_3 y^9-\frac{1}{100} x_1 x_3 y^5+\\
+\frac{1}{35}x_1 x_3 y^3-\frac{239}{12600}x_1 x_3 y.
\end{gather*}

\subsection{Mixed boundary value problem of Dirichlet-Neumann}

Solving equation (15) with the boundary conditions (22) and (23), we obtain, respectively
\begin{gather*}
K_n(|x|,y)=\mathscr{F}_t^{-1}\left[\frac{\cosh(|t|(a-y))}{\cosh(|t|a)}\right](x,y),\\
L_n(|x|,y)=\mathscr{F}_t^{-1}\left[\frac{\sinh(|t|y)}{|t|\cosh(|t|a)}\right](x,y),
\end{gather*}
and the solution of the problem is written in the form of a convolution (19), where $\phi_1 (x),\psi_1 (x)$ are polynomials. This solution is also a polynomial.

\subsubsection{The case $n = 1$.}
 Just as in the case of the Dirichlet problem,  the unique solution of a mixed boundary-value problem, defined by convolution (19) is a polynomial. We have
\begin{gather*}
u_k(x,y)=x^k*K_1(x,y)=\sum_{m=0}^{[k/2]}C_k^{2m}x^{k-2m}p_{2m}(y),\\
v_l(x,y)=x^l*L_1(x,y)=\sum_{m=0}^{[l/2]}C_l^{2m}x^{l-2m}q_{2m}(y),
\end{gather*}
where $p_{2m}(y)$ and $q_{2m}(y)$ are polynomials in $y$ and generating functions of these polinomials are
\begin{gather*}
\frac{\cosh(t(a-y))}{\cosh(ta)}=\sum_{m=0}^{\infty}p_{2m}(y)\frac{(-1)^mt^{2m}}{(2m)!},\\
\frac{\sinh(ty)}{t\cosh(ta)}=\sum_{m=0}^{\infty}q_{2m}(y)\frac{(-1)^mt^{2m}}{(2m)!}.
\end{gather*}
We give some first polynomials and the corresponding solutions of the boundary-value problem
\begin{gather*}
p_0 (y)=1,\\
p_2 (y)=-y^2+2ay,\\
p_4 (y)=y^4-4ay^3+8a^3 y,\\
p_6 (y)=-y^6+6ay^5-40a^3 y^3+96a^5 y,\\
p_8 (y)=y^8-8ay^7+112a^3 y^5-896a^5 y^3+2176a^7 y,\\
p_{10} (y)=-y^{10}+10ay^9-240a^3 y^7+4032a^5 y^5-32640a^7 y^3+79360a^9 y.\\
u_0 (x,y)=1,\\
u_1 (x,y)=x,\\
u_2 (x,y)=x^2-y^2+2ay,\\
u_3 (x,y)=x^3-3xy^2+6axy,\\
u_4 (x,y)=x^4-6x^2 y^2+12ax^2 y+y^4-4ay^3+8a^3 y,\\
u_5 (x,y)=x^5-10x^3 y^2+20ax^3 y+5xy^4-20axy^3+40a^3 xy.\\
q_0 (y)=y,\\
q_2 (y)=-\frac{1}{3} y^3+a^2 y,\\
q_4 (y)=\frac{1}{5} y^5-2a^2 y^3+5a^4 y,\\
q_6 (y)=-\frac{1}{7} y^7+3a^2 y^5-25a^4 y^3+61a^6 y,\\
q_8 (y)=\frac{1}{9} y^9-4a^2 y^7+70a^4 y^5-\frac{1708}{3} a^6 y^3+1385a^8 y,\\
q_{10} (y)=-\frac{1}{11} y^{11}+5a^2 y^9-150a^4 y^7+2562a^6 y^5-20775a^8 y^3+50521a^{10} y.\\
v_0 (x,y)=y,\\
v_1 (x,y)=xy,\\
v_2 (x,y)=x^2 y-\frac{1}{3} y^3+a^2 y,\\
v_3 (x,y)=x^3 y-xy^3+3a^2 xy,\\
v_4 (x,y)=x^4 y-2x^2 y^3+6a^2 x^2 y+\frac{1}{5} y^5-2a^2 y^3+5a^4 y,\\
v_5 (x,y)=x^5 y-\frac{10}{3} x^3 y^3+10a^2 x^3 y+xy^5-10a^2 xy^3+25a^4 xy.
\end{gather*}

\subsubsection{The case $n> 1$.}
In this case we obtain formulas analogous to the formulas given for the solution of the Dirichlet problem.

\textbf{Example 3. }
Consider a mixed Dirichlet-Neumann boundary value problem for the Poisson equation
\begin{gather*}
\Delta u(x,y)=x^{(3,2,1)}y^3=x_1^3 x_2^2 x_3 y^3,~~x\in\mathbb{R}^3,~~0<y<1,\\
u(x,0)=0,~~u_y(x,1)=0,~~x\in\mathbb{R}^3.
\end{gather*}
A particular solution of the Poisson equation is the function
\[
\tilde u(x,y)=\frac{1}{20}x_1^3x_2^2x_3y^5-\frac{1}{420}x_1^3x_3y^7-\frac{1}{140}x_1x_2^2x_3y^7+\frac{1}{2520}x_1x_3y^9.
\]
and the Dirichlet-Neumann problem for the Poisson equation reduces to the Dirichlet-Neumann problem for the Laplace equation for the function $v(x,y)=u(x,y) –\tilde u(x,y)$:
\begin{gather*}
\Delta v(x,y)=0,~~x\in\mathbb{R}^3,~~0<y<1,\\
v(x,0)=u(x,0)-\tilde u(x,0)=0,\\
v(x,1)=u_y(x,1)-\tilde u_y(x,1)=\\
=-\frac{1}{4} x_1^3 x_2^2 x_3+\frac{1}{60} x_1^3 x_3+\frac{1}{20} x_1 x_2^2 x_3-\frac{1}{280}x_1 x_3.
\end{gather*}
The solution of the Dirichlet-Neumann problem for the Poisson equation is the function
\begin{gather*}
u(x,y)=\tilde u(x,y)+v(x,y)=\tilde u(x,y)-\\
-\frac{1}{4}v_{(3,2,1)}(x,y)+\frac{1}{60}v_{(3,0,1)}(x,y)+\frac{1}{20}v_{(1,2,1)} (x,y)-\frac{1}{280}v_{(1,0,1)}(x,y)=\\
=\frac{1}{20} x_1^3 x_2^2 x_3 y^5-\frac{1}{4}x_1^3 x_2^2 x_3 y
-\frac{1}{420} x_1^3 x_3 y^7+\frac{1}{12} x_1^3 x_3 y^3-\frac{7}{30} x_1^3 x_3 y-\\
-\frac{1}{140}x_1 x_2^2 x_3 y^7+\frac{1}{4} x_1 x_2^2 x_3 y^3-\frac{7}{10} x_1 x_2^2 x_3 y+\frac{1}{2520}x_1 x_3 y^9-\frac{1}{20} x_1 x_3 y^5+\\
+\frac{7}{15}x_1 x_3 y^3-\frac{323}{280}x_1 x_3 y.
\end{gather*}

\section*{                                             Conclusion}
    
It is shown that the Poisson equation with the polynomial right-hand side and with the polynomial boundary conditions of Dirichlet and Dirichlet-Neumann in the layer, has a unique solution in the class of functions of slow gowth and this solution is a polynomial. An algorithm for constructing this polynomial solution is given. Examples are considered.


\begin{thebibliography}{99}
\bibitem{S}
Sobolev S.L. Partial differential equations of mathematical physics. Pergamon Press , 1964.
\bibitem{SW}
Stein E., Weis G. Introduction to Fourier Analysis on Euclidean Spaces. Princeton University Press, 1971. 296 p.
\bibitem{N1}
Nikol'skii S.M. A boundary value problem for polynomials //Proceedings of the Steklov Institute of Mathematics, 1999,  227, pp. 223-236. 
\bibitem{N2}
Nikol'skii S.M. More about the boundary-value problem with polynomials // Proceedings of the Steklov Institute of Mathematics, 2001,  232, pp. 286-288.
\bibitem{K}
Karachik V. V. Construction of polynomial solutions to the Dirichlet problem for the polyharmonic equation in a ball //Computational Mathematics and Mathematicals Physics, 2014,  54, No. 7, pp. 1122-1143.
\bibitem{V1}
Volkov E.A. A criterion for the solvability of boundary value problems for the Laplace and Poisson equations on special triangles and a rectangle in algebraic polynomials // Proceedings of the Steklov Institute of Mathematics, 1999,  227, pp. 122-136.
\bibitem{V2}
Volkov E.A. On the solvability in the class of polynomials of the Dirichlet problem for the Laplace equation on an arbitrary polygon // Proceedings of the Steklov Institute of Mathematics, 2001,  232, pp. 102-114.
\bibitem{AK1}
Algazin O.D. , Kopaev A.V., Solution to the Mixed Boundary-Value Problem for Laplace Equation in Multidimensional Infinity Layer,  Herald of the BMSTU, Series “Natural Sciences” (2015), No.1, 3-13. 
    DOI:10.18698/1812-3368-2015-1-3-13.

\bibitem{AK2}
Algazin O.D. , Kopaev A.V., Solution of the Dirichlet Problem for the        Poisson’s equation in a Multidimensional Infinity Layer, Mathematics and Mathematical Modelling of the Bauman MSTU (2015), No.4, 41-53 .
    DOI: 10.7463/mathm.0415.0812943.

\bibitem{V}
Vladimirov V.S. Generalized functions in mathematical physics. Moscow, Nauka Publ., 1979.320 p.
\bibitem{P}
Polyanin A.D. Handbook of linear equations of mathematical physics. Moscow, Fizmatlit Publ., 2001. 576 p.
\bibitem{W}
Widder D.V. Functions harmonic in a strip//Proc. Amer. Math. Soc. 1961. pp.67-72.
\bibitem{B}
Brawn F.T. The Green and Poisson kernels for the strip $\mathbb{R}^n\times]0,1[$ //J. London Math. Soc.(2),2. 1970. pp.439-454.
\bibitem{PBM}
Prudnikov A.P., Brychkov Yu.A. Marichev O.I. Integrals and series. Elementary functions. Moscow, Nauka Publ., 1981. 800 p.


\end{thebibliography}
\end{document}